\documentclass[letterpaper,twocolumn,10pt]{article}
\usepackage{usenix,epsfig,endnotes}
\usepackage[dvipsnames,usenames]{color}
\usepackage{caption}
\usepackage{subcaption}
\usepackage{fancyvrb}
\usepackage{setspace}
\usepackage{url}
\usepackage{color}
\usepackage{paralist}
\usepackage[small,compact]{titlesec}
\usepackage{listings}
\usepackage{ifthen}

\newcommand{\DONE}[1]{}
\newcommand{\COMMENT}[1]{}

\newcommand{\figref}[1]{Fig.~\ref{Fi:#1}}

\newcommand{\secref}[1]{Section~\ref{Se:#1}}

\newcommand{\ssecref}[1]{Sec.~\ref{Se:#1}}

\newcommand{\figlabel}[1]{\label{Fi:#1}}

\newcommand{\seclabel}[1]{\label{Se:#1}}
\newcommand{\sseclabel}[1]{\label{Se:#1}}

\newcommand{\ignore}[1]{}

\newboolean{TR}
\setboolean{TR}{false}
\ifthenelse{\boolean{TR}}{

\newcommand{\TrOnly}[1]{#1}
\newcommand{\SubOnly}[1]{}
\newcommand{\TrOnlyInFootnote}[1]{#1}
\newcommand{\TrOnlyInTable}[1]{#1}}
{

\newcommand{\TrOnly}[1]{}
\newcommand{\SubOnly}[1]{#1}
\newcommand{\TrOnlyInFootnote}[1]{}
\newcommand{\TrOnlyInTable}[1]{}}

\newcommand{\hiddentext}[1]{}

\newcommand{\para}[1]{\vspace{3pt}\noindent\textbf{\textit{#1}}}

\newcommand{\sname}[1]{{\small \textsc{#1}}}

\renewcommand{\phi}{\varphi}

\begin{document}

\date{}

\title{Exploiting Social Navigation}

\author{
{\rm Meital Ben Sinai} \\ Technion, Israel
\and {\rm Nimrod Partush} \\ Technion, Israel
\and {\rm Shir Yadid} \\ Technion, Israel
\and {\rm Eran Yahav}\\ Technion, Israel
} 
\maketitle

\thispagestyle{empty}

\begin{abstract}
We present an effective Sybil attack~\cite{DBLP:conf/iptps/Douceur02} against social location based services. Our attack is based on creating a large number of reputed ``bot drivers'', and controlling their reported locations using fake GPS reports. We show how this attack can be used to influence social navigation systems by applying it to \sname{Waze}~\cite{Waze} - a prominent social navigation application used by over $50$ million drivers. We show that our attack can fake traffic jams and dramatically influence routing decisions. We present several techniques for preventing the attack, and show that effective mitigation likely requires the use of additional carrier information.
\end{abstract}

\section{Introduction}

Social navigation is rapidly becoming a prevalent approach for navigation. A social navigation system collects navigation data such as route duration, traffic congestion, road obstacles, and even map layout from users, and then uses this information to route other users.

\sname{Waze} is a prominent social navigation system used by over $50$ million users worldwide (recently acquired by Google). At this scale, social navigation data is considered a reliable source of traffic information trusted by many. For example, \sname{Waze} traffic information is used to feed reports for various Radio and TV stations~\cite{WazeOnTV}, as well as the widely used Google Maps navigation application~\cite{GoogleMapsWaze}. This propagates the influence of social navigation to a much wider community.

In this paper, we present the first Sybil attack on social navigation systems, and demonstrate its effectiveness by applying it to \sname{Waze}. In a Sybil attack the attacker subverts the reputation system of a network by creating a large number of pseudonymous identities, using them to gain a disproportionately large influence~\cite{DBLP:conf/iptps/Douceur02}. Our main idea is to create a large number of reputed ``bot drivers'' and use fake GPS information to drive them in desired traffic patterns. For example, the ``bots'' can be used to simulate patterns detected by \sname{Waze} as ``heavy traffic''.

In contrast to previous attacks on \sname{Waze}~\cite{FloatingCar}, our approach does not require reverse engineering the client, or the protocol, and cannot be mitigated by hardening the protocol as it uses legitimate clients.

We demonstrate the effectiveness of our attack by applying it to \sname{Waze}, and showing that  our technique can simulate traffic jams, and dramatically influence the routing decisions presented to users.

The attack has vast security and financial implications. For example, the attack could be targeted towards specific businesses or toll roads, falsely reporting the area as congested and discouraging drivers from nearing it. The attack also affects drivers safety, where drivers may be sent along a less suited road, or distracted by spurious obstacle reports. We also discuss privacy issues due to disclosure of user information on the \sname{Waze} ``live map''.

\paragraph{Main Contributions} The main contributions of this paper are:
\begin{compactitem}
\item We present a Sybil attack on social navigation systems.
\item Our attack is cheap and can be facilitated using free off-the-shelf emulation software, a simple fake GPS player application, controlled by the Android Debug Bridge, running on a 16 core machine.
\item We demonstrate our approach by simulating traffic jams, and affecting the routing presented to drivers on the \sname{Waze} social navigation system.
\item We present two approaches for mitigating our attack and similar attacks on social navigation: using carrier information and user behavioral analysis. We evaluate them based on parameters of simplicity, user experience, effectiveness and cost.
\end{compactitem}

\section{Attacks} \seclabel{Attacks}
In this section we describe all successfully implemented attacks. We first describe how to create bot drivers and increase their reputation, and then describe how bot drivers are used for various attacks.

\subsection{Creating Bot Drivers} \seclabel{CreateBotDrivers}
\para{Becoming an influential part of the \sname{Waze} community requires a single click.} Just by installing and starting the application on her device, a user is able to view surrounding user and traffic information, influence nearby users by reporting various road obstacles, and affect routing and traffic information. This allowed us for automated creation of bot drivers simply by starting an emulator, running \sname{Waze}, and running a script that clicks the ``start driving'' button.

\para{Registration does not require validation} \sname{Waze} offers a registration process, encouraged by a rating system where users receive points for using the application and reporting obstacles. The registration process however, employs no validation mechanisms. We assume \sname{Waze} intentionally avoids user validation, in hopes of encouraging new users to try the application. Interestingly, deleting a \sname{Waze} account does require the user to pass a {\small CAPTCHA}. This further establishes our assumption that the \sname{Waze} registration process was designed to gather and maintain users, and not to validate them.

\para{Training Reputed Bots} Due to lack of validation, we were able to automate the process of \emph{creating a registered \sname{Waze} user}. We used random bot usernames to avoid clashes with pre-existing names. Registration does require a well-formed email address, but it is not further verified. Throughout our experiments, we re-used the same user accounts and were able to gain rating, by simulating driving and reporting obstacles. The ability to automatically register and gain rating, makes it harder to mitigate attacks by relying on user rating and tenure. In fact, by emulating benign user behavior, attempts to eliminate bots on behavioral basis are likely to fail.

%

\subsection{Creating Traffic Jams and Influencing Routing} \seclabel{CreateTrafficJam}
Next, we describe how we were able to simulate traffic jams and influence the \sname{Waze} routing algorithm. \sname{Waze} deduces traffic congestion and routing time information from location and movement data reported by its users~\cite{WazeRoutes}. This algorithm resides on the server side of the \sname{Waze} system and was never publicly disclosed. The main challenge of this work, was in fact experimentally deducing and exploiting this algorithm.

Our experiments consisted of explorative adjustment of the following parameters:
\begin{inparaenum}[(i)]
\item Data set size (number of bots),
\item Drive duration,
\item Speed and movement pattern.
\end{inparaenum}
All parameters were aimed at matching the \sname{Waze} traffic jam scenario. We show that with a small amount of resources, we were able to simulate a traffic jam.

\para{Gaining reputation} The \sname{Waze} application states: \emph{``As you advance to higher levels your reports get greater influence''}. We therefore trained our bot drivers by driving on campus in average speeds for several hours.

\para{Simulating standstill traffic} Our initial attempts at simulating congestion consisted of spawning botnets (groups of ``bot drivers'') of increasing sizes ($5$, $10$, $15$ etc.) and statically placing them at the target location. We assumed \sname{Waze} would acknowledge these static drivers as stuck in traffic, and report a traffic jam. This approach did not work, even after scattering the bots along the target road in different patterns, attempting to mimic a more realistic standstill traffic scenario. We attribute this result to \sname{Waze} interpreting the scenario as users entering their vehicle and starting \sname{Waze}, but have yet to move - a common scenario among navigation users.

\para{Simulating slowdowns} Our next round of experiments consisted of simulating a gradual slowdown in traffic. As before, we sent increasingly larger groups of bots to the target location, but this time they moved through the area in gradually slower speeds of $15$, $8$ and $2$ kph. At first, each bot passed through the area once for each speed in the set, then twice, etc., up to the point where each bot spends $20$ minutes passing the route (again and again) at a certain speed before moving on to the next slower speed. Lastly, we incorporated short periods of standing still (10 seconds) into each pass, also in increasing durations. This scenario still failed to yield congestion, as it required one final addition, described next.

\begin{figure}[h]
\centering
\includegraphics[scale=0.45,clip=true,trim = 10pt 0pt 0pt 0pt]{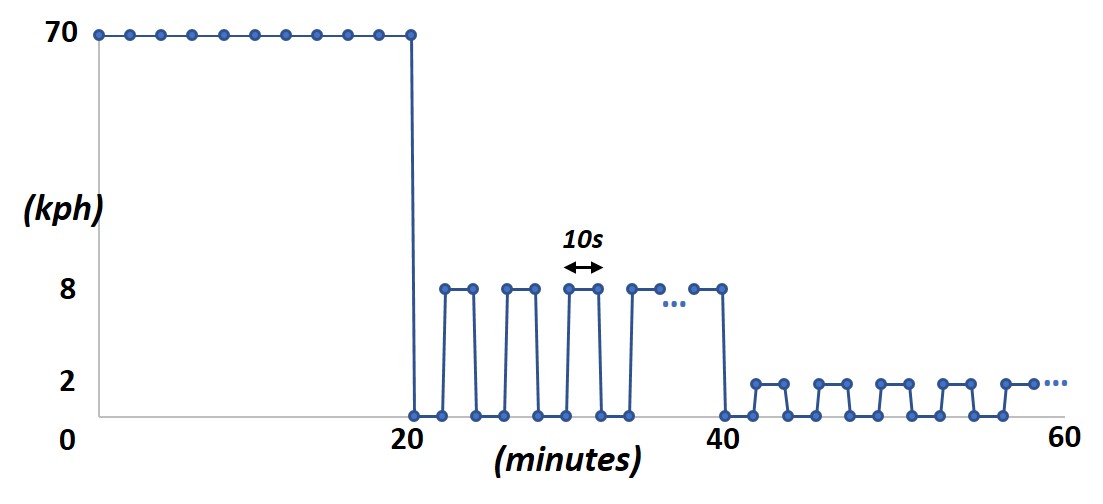}
\caption{Speed pattern of bots participating in the attack}
\figlabel{SpeedGraph}
\end{figure}

\para{Speed is relative.} We conducted our experiments on the closed campus area (see \ssecref{ExpLocation}). Initially, we saw that even when consistently simulating gradually slowing traffic, with large botnets of reputed drivers, no congestion was reported. One way this result could be explained is that \sname{Waze} takes other, non user-related parameters into effect. These parameters could include: time of day, date, weather conditions and \emph{road information}. We in fact learned, that since the campus area is a very low speed zone, \sname{Waze} regarded all slow driving traffic through it as normal rather than congested. In fact, it seems that the \sname{Waze} congestion reporting algorithm is a relative one, i.e., a route is congested if its current average speed is considerably lower \emph{relative to former known speeds}. Thus, we added another phase to the bots driving pattern, where driving was initiated at 70kph. After a period of $20$ minutes, we reverted to the previous pattern of simulating slowdown in traffic, finally leading to successful simulation of a traffic jam. \figref{SpeedGraph} shows the speed over time of each bot in the successful implementation of the attack. The successful attack consisted of only $15$ bots, driving at the pattern shown in \figref{SpeedGraph}, where we reset a bot location to the route origin whenever it has reached the end of the route (and thus enable it to drive the route again according to the desired speed pattern).

\para{Simulated traffic jams on the \sname{Waze} map.} \figref{TrafficJam} shows a traffic jam created by our attack. The congestion is marked by a bold red line, annotated with arrows showing the congested direction, along with a notification marking the average speed as 8kph. The congested route also features the botnet used for the attack, showing as group of \sname{Waze} users along the start of congested route (we further discuss users shown on the live map in~\ssecref{TrackingUsers}). We observed that traffic jams created by the attack persists for roughly $20$ minutes past the point where the bots stop driving.

\begin{figure}[h]
\centering
\includegraphics[width=0.85\linewidth,clip=true,trim = 190pt 50pt 195pt 40pt]{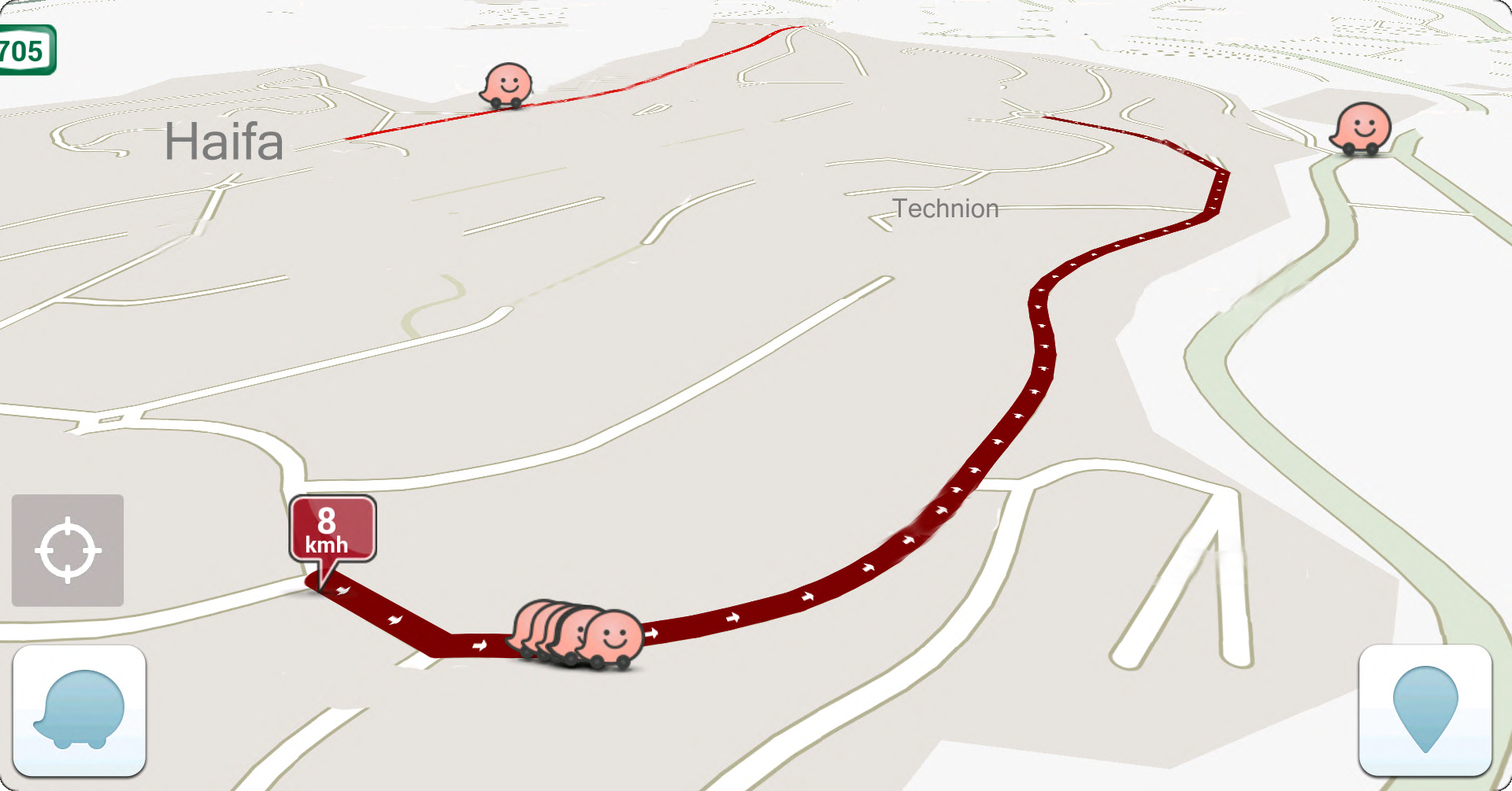}
\caption[]{Reported traffic jam, created by the attack\footnotemark}
\figlabel{TrafficJam}
\end{figure}

\footnotetext{\sname{Waze} map images included Hebrew and thus were edited for readability. Link to original images: \url{http://www.cs.technion.ac.il/~nimi/waze.html}}

\para{Influencing routing} We were able to influence the \sname{Waze} routing algorithm and potentially send benign users on different routes. This has vast security and financial implications. The attack can be used to ease actual congestion in roads of interest to the attacker, by falsely reporting them as congested, and sending away benign users. Alternately, an attacker may congest areas causing potential financial damage, due to time and fuel wasted by drivers taking longer routes, or drivers avoiding using tool roads or visiting businesses residing in congested areas. This also a safety issue, where drivers may be sent on a less suited road. Furthermore, using the attack, one is able to completely determine the route planned by \sname{Waze} for a user trying to get to a destination. Consequently, users will be directed along a desired path controlled by a malicious attacker. \figref{RoutePreJam} shows the recommended route, prior to an attack. The route is marked by a bold purple line, annotated with street names, and reported to be 1.5km long with 3 minutes traversal time. \figref{RoutePostJam} shows the change in route due to the fake traffic jam we created along the previous route. The new route is 1.6km long and takes 4 minutes to traverse.

\begin{figure}[h]
\centering
\begin{subfigure}{.5\linewidth}
\centering
\includegraphics[clip=true,trim=170pt 0pt 110pt 350pt,width=.93\linewidth]{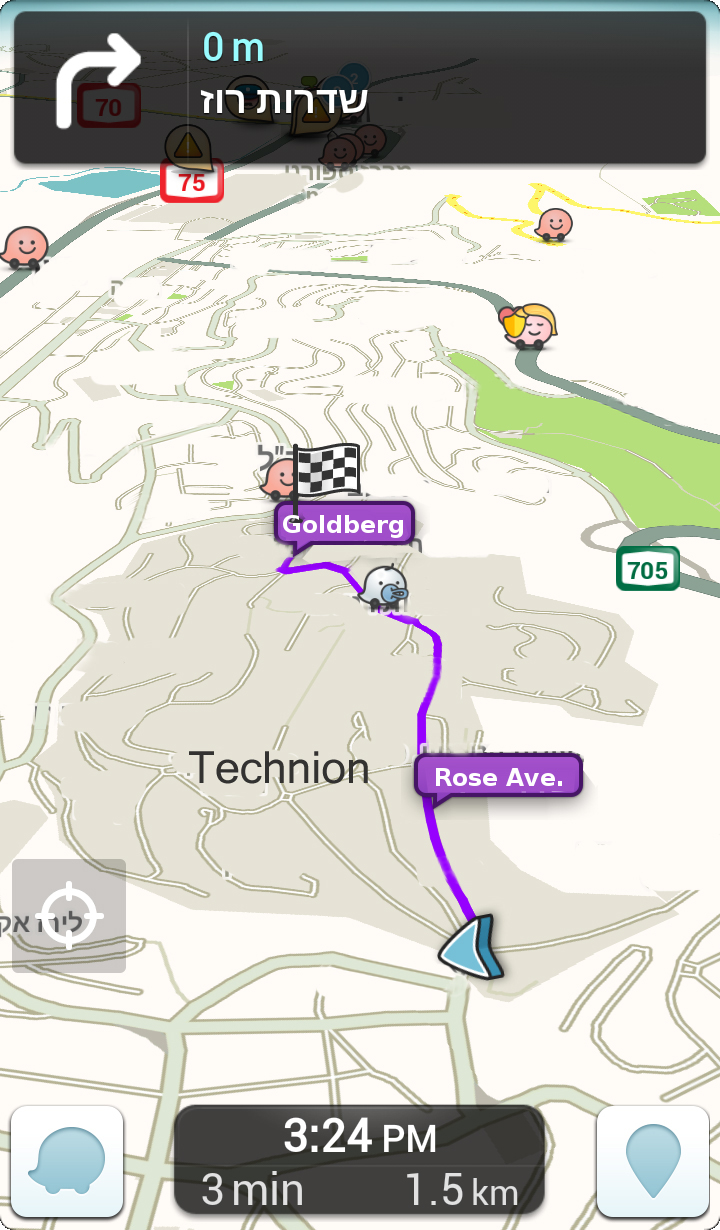}
\caption{Pre-attack route}
\figlabel{RoutePreJam}
\end{subfigure}%
\begin{subfigure}{.5\linewidth}
\centering
\includegraphics[clip=true,trim=110pt 0pt 170pt 350pt,width=.93\linewidth]{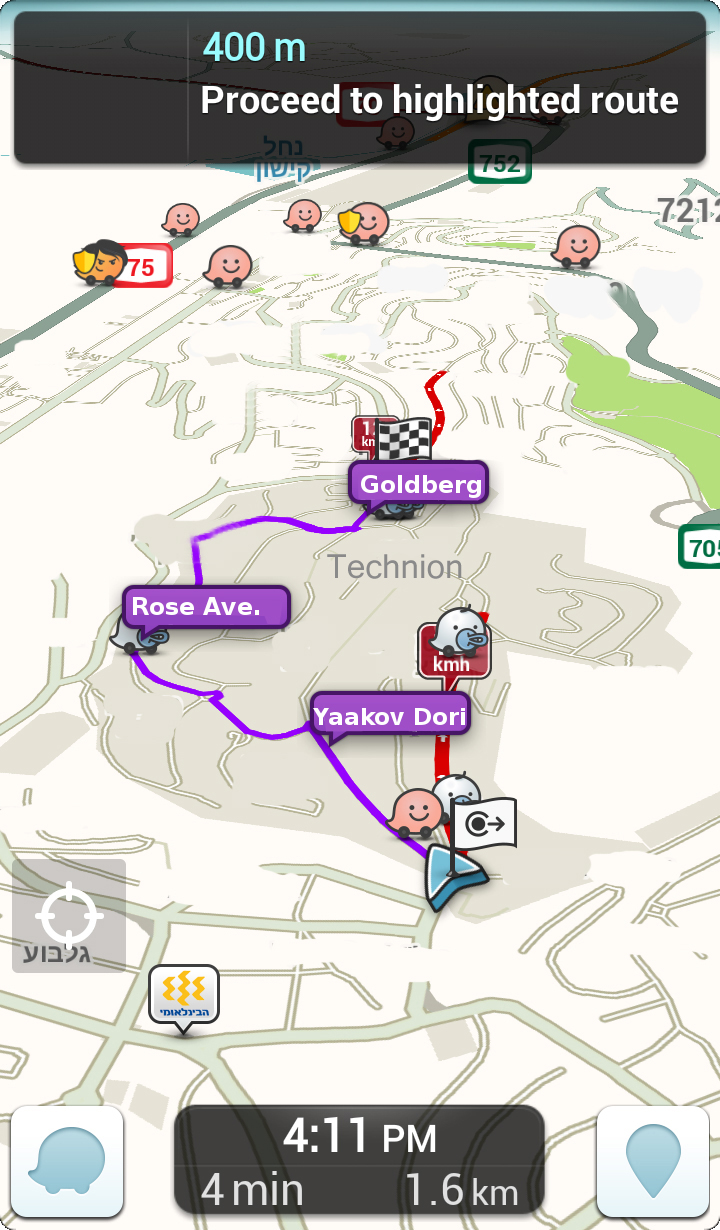}
\caption{Post-attack route}
\figlabel{RoutePostJam}
\end{subfigure}
\caption{\sname{Waze} routing, influenced by the attack}
\end{figure}


\para{Thwarting machine-like behaviour} Throughout our experiments, we explored how \sname{Waze} reacts to machine-like behavior, such as spawning all bots at the exact same time or driving at the same speed, along the same route, all at once. We spent a significant amount of time planning our experiments to simulate benign human behavior. We added a mechanism for gradual creation of the driver botnets, with random time spacing between each spawn. We also implemented our command and control interface such that each bot could receive a different movement pattern, along with randomization. After successfully implementing the attack, we proceeded to verify the necessity of these mechanisms. We retried the traffic jam scenario, while disabling each of these mechanisms, and learned that a congestion was reported regardless. This means that \sname{Waze} makes no observable effort to differentiate human from machine behavior. Ultimately, this may have been a wise design choice, as applying such mechanisms is destined to fail against a highly skilled attacker. However, in \secref{MitigatingAttacks} we describe two techniques that can mitigate such attacks.

\subsection{Reporting Obstacles} \seclabel{ReportingObstacles}

\begin{figure}[h]
\centering
\includegraphics[width=0.7\linewidth,clip=true,trim = 0pt 0pt 0pt 0pt]{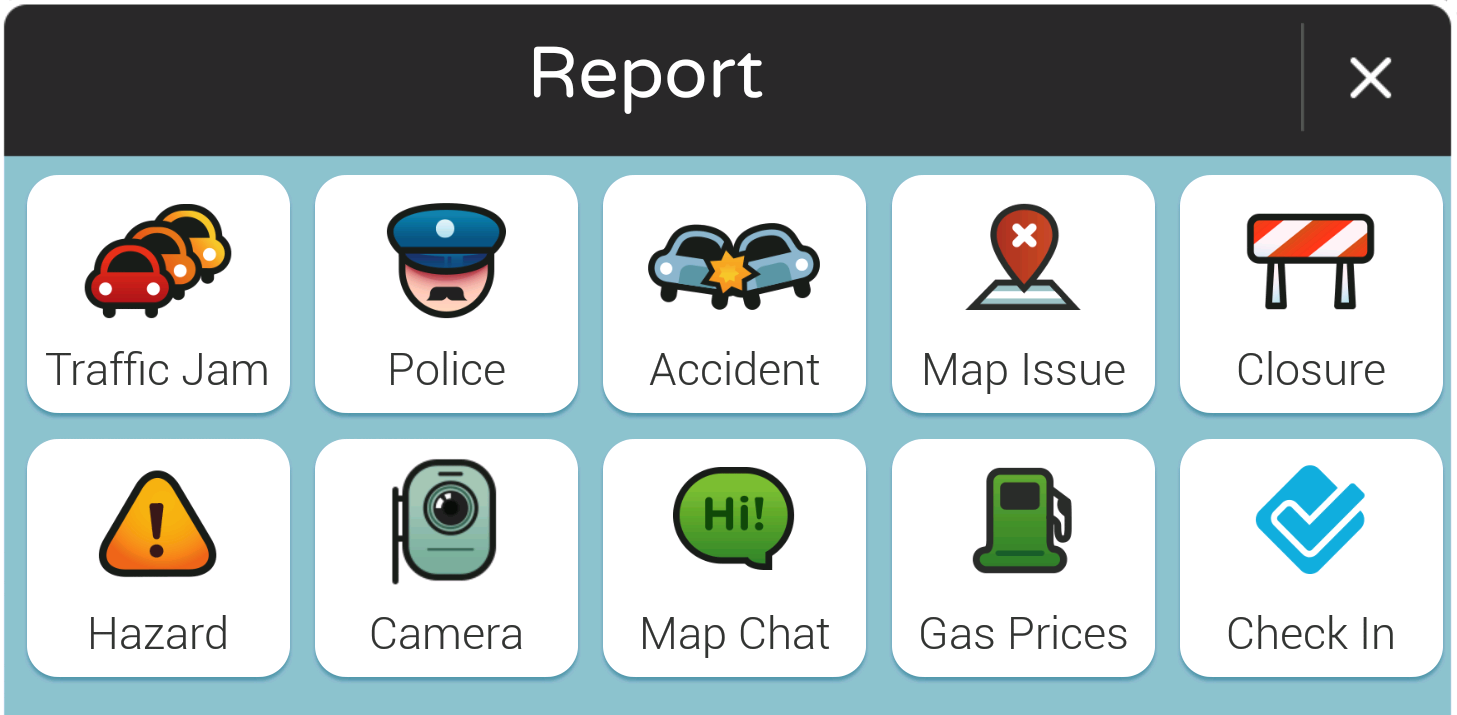}
\caption{\sname{Waze} obstacle reporting menu}
\figlabel{ReportMenu}
\end{figure}

\para{Reporting persistent spurious obstacles} The \sname{Waze} application allows notifying nearby users of the presence of obstacles on the road. \figref{ReportMenu} depicts the \sname{Waze} obstacle reporting interface. Reporting obstacles can be done by any user, whether registered or not. The reporting process consists of a few clicks, and users are also able to add a note or take a picture of the reported obstacles. Once the report is complete, the obstacle immediately appears on the map of all nearby users, and stays there for roughly 20 minutes. The duration can be extended, however, if reported as credible by other users. Thus, fake sustainable reports can be easily produced by a small group of bot drivers, one reporting an obstacle with a designated note and the rest confirming the report. We note that reporting a ``Traffic Jam'' obstacle does not effectively create a traffic jam as in \figref{TrafficJam}, or affect routing. Instead, a notification is shown on the map, left to the discretion of the user.

\para{Indirectly influencing traffic} Although obstacle reports do not influence routing (\sname{Waze} will not prefer a non-obstructed road over an obstructed one), they have a strong mental effect on human users. A user observing multiple obstacles along a certain route, some containing notes supposedly written by other users, may prefer taking a different route, or drive much slower through the obstructed route. This could have an indirect effect on traffic congestion and can be performed with a minimal amount of resources.

\para{Report system DDos} We also considered attacks where the report mechanism is rendered useless. This can be performed by blasting the application with a huge amount of spurious reports, in the form of a DDos attack. \figref{PoliceObstacles} shows the effect of reporting multiple police units and hazards over a small region. \sname{Waze} limits the number of reports one user can produce in a short time frame, but this limitation can be easily overcome by closing the application and logging as a different user (reports remain after the reporting user closed the application).

\begin{figure}[h]
\centering
\includegraphics[width=0.85\linewidth,clip=true,trim = 200pt 100pt 200pt 0pt]{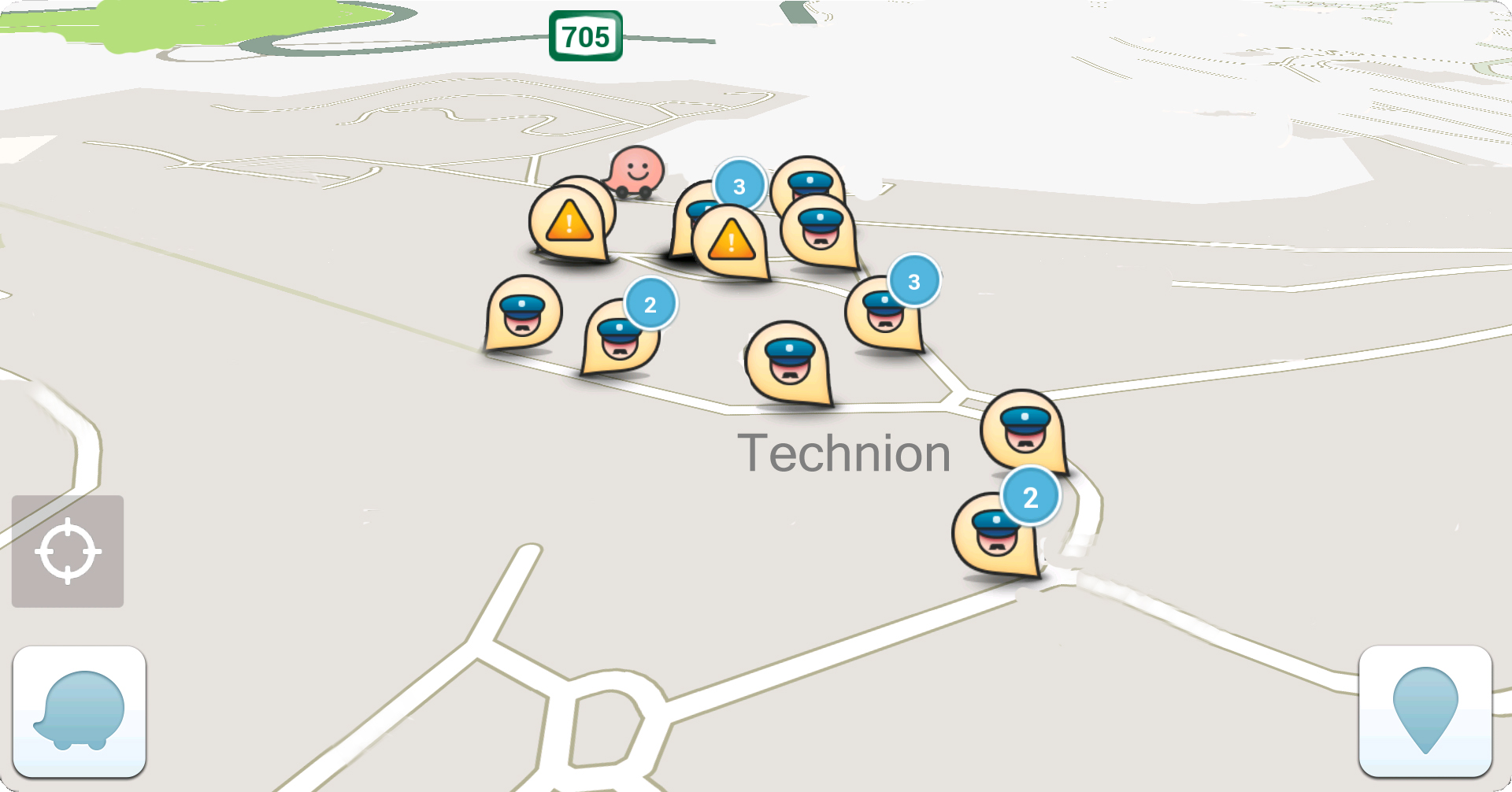}
\caption{Fake reports of police units on the \sname{Waze} map}
\figlabel{PoliceObstacles}
\end{figure}

\para{Invalidating benign reports} Another similar attack can be performed by invalidating benign user reports. Whenever a \sname{Waze} user approaches an obstacle, they are given the option to update the report as outdated. An attack where the \sname{Waze} map is constantly combed for reports and those are, in turn, invalidated by a bot driver, will incapacitate the report system and in fact give the attacker full control of said system.

We note that the \sname{Waze} report system also allows directly influencing map layout, by reporting road closures and map issues. However, we refrained from exploring such scenarios as the risk of damaging benign \sname{Waze} users was too great. Should this attack be made possible, financial and safety implications are dire, where users may completely avoid a closed road or be sent along a non-existing path.

\subsection{Tracking Users} \seclabel{TrackingUsers}

\begin{figure}[h]
\centering
\includegraphics[width=0.85\linewidth,clip=true,trim = 200pt 0pt 200pt 150pt]{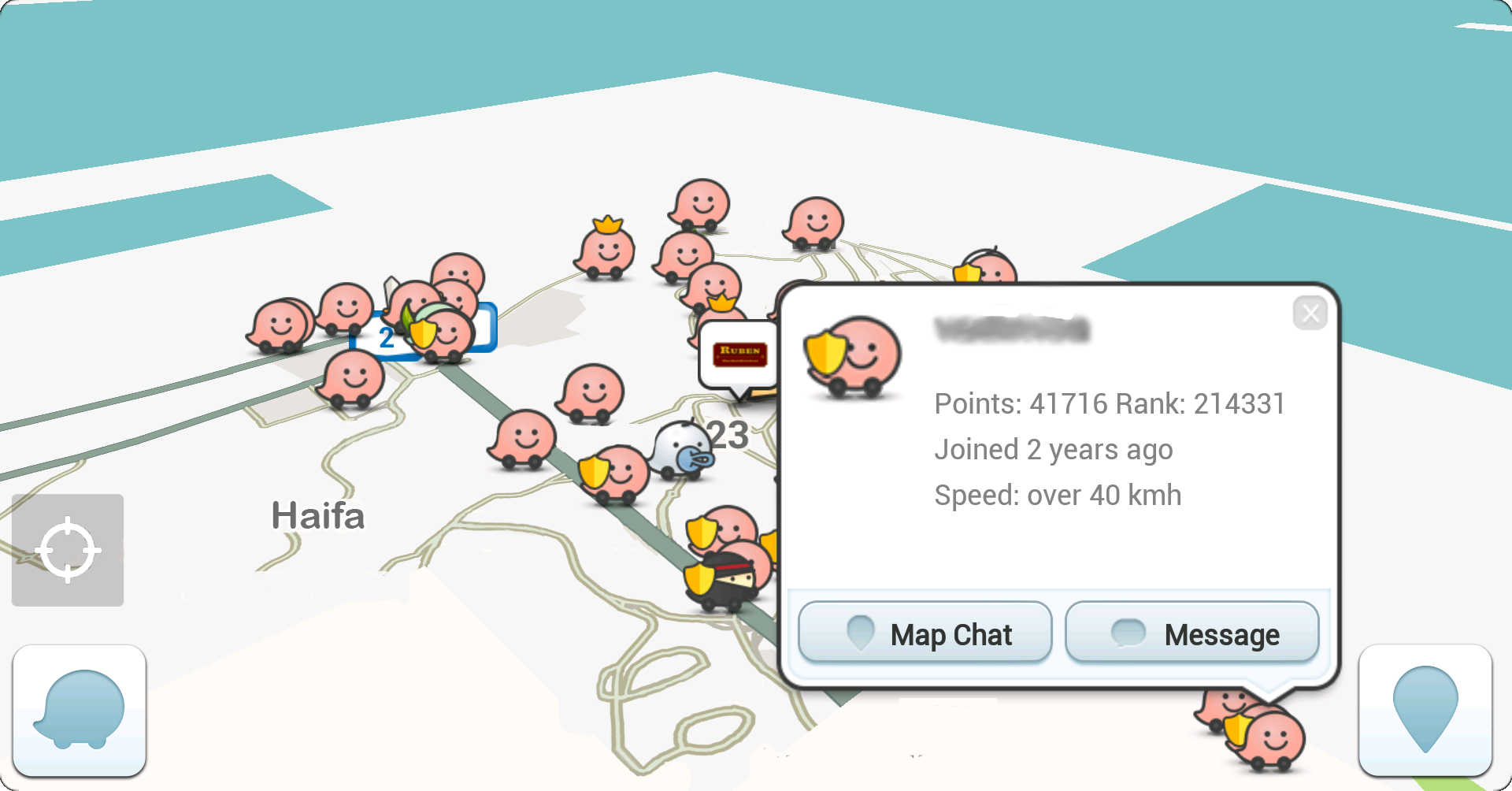}
\caption{\sname{Waze} live map}
\figlabel{LiveMap}
\end{figure}

\para{Statistically tracking users via the live map} \figref{LiveMap} shows the \sname{Waze} ``live map'', which functions as the main view of the application. The live map features details of fellow \sname{Waze} users, over all locations, randomly appearing in and out of the map. Further, one can receive detailed information on a fellow user by clicking on their icon, as seen in \figref{LiveMap} (blurred for privacy reasons). It is also possible to chat with or message the selected user, which adds social network-like features to \sname{Waze}. Revealed user information includes username, points and ranking, seniority and speed. This information may be leveraged towards compromising user privacy. An attacker aiming to track a certain username, could use this data to their advantage. We did not fully explore this attack, however our experiments suggest that (i) live maps differ from one user to another, i.e. each user receives a list of different usernames (ii) all users appear on some user's live map, at some point. These findings can be used to perform a statistical attack, where live maps will be constantly scanned, searching for a target username. Since each client views a different set of users, with a sufficient amount of clients, the whereabouts of a targeted user could be ascertained. Gathered locations could be grouped together into a partial route to be used for tracking unsuspecting users. We observed that user locations on the live map are precise up to a time frame of 5 minutes. i.e. if a user appears on the live map at a certain location, they were at that location within the past 5 minutes. User speed however, was observed to be inaccurate.





\section{Experiment System Description}
In this section we will shortly describe the experiment parameters and tools developed to carry out attacks.

\subsection{Emulating \sname{Waze} Clients}
For the experiment purposes, we required a non-costly, easily deployed system that will allow us to create and control a large amount of \sname{Waze} clients. This suggested a software emulation solution and after some research we arrived at using Android Developer Tools software emulator. This solution complied with our requirements as it allowed us to run multiple instances over our Linux servers. We used the Android Developer Tools version 22.3.0 with Android API 17 and ran the \sname{Waze} Android client version 3.5.1.4.

\subsection{Command and Control Interface}
To allow for rapid development, we developed our system in Python. We controlled the android clients through the Android Debug Bridge (adb), a versatile command line tool that allows communicating with an emulator instance. After starting each emulator instance, \sname{Waze} was installed on it through a series of ADB commands, simulating text and touch input. In order to successfully automate controlling the \sname{Waze} client on the emulator instance, we used the \sname{Waze} application activity log, as it was updated with every completed operation (we assume for debugging). We suggest removing said logging from future versions of \sname{Waze}.

\subsection{GPS Data}
In order to emulate driving along a route, we constructed a small Android application which generates mock GPS locations adhering to driving on the target route, according to a given movement pattern. Our application, appropriately named \texttt{TrafficJam}, receives a start point, end point and direction and then proceeds to generate mock GPS locations along that route. \sname{Waze} considers these mock GPS locations to be valid and moves the client accordingly on the map. The application also receives a movement pattern, composed of an average speed, duration of movement and standstill intervals, and generates the mock GPS locations accordingly, emulating moving along the desired route, at the given speed, for the given time duration, while standing still at certain intervals.

\figref{CreateTrafficJamCode} contains pseudo-code of the congestion creation script, implementing the attack described in \secref{CreateTrafficJam}.

\newcommand\funcHighlight[1]{\textcolor[rgb]{1,0,0}{\textbf{#1}}}
\newcommand\textHighlight[1]{\textcolor[rgb]{1,0.5,1}{\textbf{#1}}}
\newcommand\unitHighlight[1]{\textcolor{blue}{\textbf{#1}}}
\begin{figure}[h]
\begin{lstlisting}
|\funcHighlight{CreateTrafficJam}|(num_bots,route,reputed_users) {
  emulators = |\textbf{StartEmulators}|(num_bots);
  |\textbf{InstallAndStartApp}|(emulators,|\textHighlight{"Waze"}|);
  |\textbf{InstallAndStartApp}|(emulators,|\textHighlight{"TrafficJam"}|);
  |\textbf{LoginWaze}|(emulators,reputed_users);
  stops=|\textHighlight{"none"}|;
  |\textbf{TrafficJam}|(emulators,route,70|\unitHighlight{kph}|,20|\unitHighlight{min}|,stops);
  stops=|\textHighlight{"stop for 10s every 10s"}|;
  |\textbf{TrafficJam}|(emulators,route,8|\unitHighlight{kph}|,20|\unitHighlight{min}|,stops);
  |\textbf{TrafficJam}|(emulators,route,2|\unitHighlight{kph}|,20|\unitHighlight{min}|,stops);
}
\end{lstlisting}
\caption{Congestion creation script pseudo-code}
\figlabel{CreateTrafficJamCode}
\end{figure}

\subsection{Experiment Location}\seclabel{ExpLocation}
As our experiments affect real \sname{Waze} users, we avoided using major roads and highways. We ran our experiments on Technion campus in the city of Haifa, Israel. We chose this region as it is small enough and relatively closed off to have a minor affect on the general user population. However, the campus road system is large enough to allow several navigation options. More importantly, since the campus is frequented by a large and technology-friendly crowd, we had ample amount of legitimate \sname{Waze} users to establish our results as credible.

\section{Mitigating Attacks}\seclabel{MitigatingAttacks}

In this section, we will discuss several means for mitigation attacks, while comparing them by parameters of simplicity, user experience, security and cost.

\subsection{Verifying Drivers via Carrier Data} \sseclabel{CarrierData}
\para{Relying on carrier data in social networks and mobile applications} An established trend for verifying users of mobile applications and social networks is the use of mobile carrier data. Facebook~\cite{FacebookVerifyAccount}, Google,~\cite{GoogleVerifyAccount} and many prominent mobile applications such as WhatsApp~\cite{WhatsApp} and Viber~\cite{Viber}, encourage and even require their users to verify their account using a legitimate mobile phone number. While users of traditional social networks like Facebook and Google+ may not always have a mobile number, for \sname{Waze} this is not the case, as it was designed to be run on a mobile device.

\para{Linking users to a mobile carrier identity} Retrieving the user's phone number upon registration and validating it would make the attacks of~\secref{CreateBotDrivers} considerably more complex and costly. \sname{Waze} could still allow for non-verified use of the application, but should give much smaller weight to these type of users. Note that this incurs a one time carrier cost per user.

\para{Verifying user location via carrier} Mitigating attacks in \secref{CreateTrafficJam} and \secref{ReportingObstacles} requires validating the actual location of the reporting entity. This can be done by verifying user reported GPS locations with the location of the cellular antenna carrying it. To lower costs and net traffic, this validation can occur periodically, at random times selected by the application. This solution can also be used to mitigate user tracking (\secref{TrackingUsers}), by disclosing fellow user information only to verified drivers. Note that all attacks described in \secref{Attacks} would still be possible, but would require purchasing a large amount of mobile carrier equipment (SIM cards and equipment for operating them), that would then need to be located in the vicinity of the target area (to be considered as verified).

Ultimately, we consider using carrier data to be the most secure method (as it relies on external verified data), fairly simple to implement, and user friendly. The disadvantage is the possible incurred cost by the carrier.

\subsection{Verifying Drivers via Behavioral Analysis}
Distinguishing human and robot behavior in Sybil attacks has been the subject of much study~\cite{Wang:2013:YYC:2534766.2534788,DBLP:conf/nsdi/TranMLS09,DBLP:journals/ton/YuKGF08,DBLP:journals/ton/YuGKX10}. However, in the context of verifying driver behavior, the subject has remained fairly unstudied. We briefly discuss several means for mitigating attacks based on user behavior analysis.

\para{Relying on existing validation mechanisms} Like many other mobile application, \sname{Waze} offers the option to login via Facebook account. This can be further extended towards using Google and Apple accounts, or simply asking users to solve a {\small CAPTCHA}. Granting better standing to validated users could help mitigate attacks in \secref{Attacks}, but is dependant on the validity of the underlying mechanisms~\cite{FacebookFakeAccounts,DBLP:conf/uss/MotoyamaLKMVS10}.

\para{Network traffic analysis} Our attack consisted of numerous client reports originating from a small number of static IP addresses. A simple network analysis could have thwarted these reports and treated them as unreliable. This approach could be further extended so that reports originating from IP addresses affiliated with 3G carriers~\cite{DBLP:conf/imc/BalakrishnanMR09} receive better standing. Although this method is somewhat related to carrier data, we did not include it in \ssecref{CarrierData}, as it could be performed relatively unbeknownst to the carrier. A skilled attacker could subvert this defence mechanism by using multiple 3G network access, incurring extra attack costs.

\para{Analyzing user creation, movement and report patterns} In \ssecref{CreateTrafficJam} we discussed possible mechanisms of disguising bot behavior, such as gradual creation of botnets and different movement patterns for each bot. These could be mitigated by performing a more in-depth analysis of driver behavior and comparison with benign user behavior. Unfortunately, these break against a skilled opponent, successfully mimicking human drivers. The same technique could be applied to obstacle reports, preventing spurious reports and invalidation of benign reports.

The main advantage of this approach is that it could be performed entirely in-house, without involving third party components which incur extra costs and net traffic. User experience is affected by registration validation. However, as mentioned, analysing user behavior is fairly complicated and can requires constant maintenance and upgrade, in order to stay ahead of attackers.

\section{Conclusion}

We showed that a Sybil attack on a location based service is possible, and could affect large communities of users, causing financial damage and compromising security and privacy. Our attack requires a low amount of resources, and can be performed by many attackers. We offer ways for mitigating the attack, balancing simplicity, effectiveness and cost.  

{\footnotesize
\begin{spacing}{0.9}
\bibliographystyle{acm}
\bibliography{bib}
\end{spacing}
}

\end{document}